# 11,11,12,12-tetracyanonaphtho-2,6-quinodimethane in Contact with Ferromagnetic Electrodes for Organic Spintronics

Shengwei Shi[1,3,*], Zhengyi Sun[2], Xianjie Liu[3], Amilcar Bedoya-Pinto[4], Patrizio Graziosi[5], Huangzhong Yu[6], Wenting Li[1], Gang Liu[1], Luis Hueso[4], Valentin A. Dediu[5], and Mats Fahlman[3,*]

[1]*Hubei Key Laboratory of Plasma Chemistry and Advanced Materials, School of Materials Science and Engineering, Wuhan Institute of Technology, 430205 Wuhan, China.* E-mail: shisw2001@hotmail.com

[2]*Institute of Advanced Materials, Jiangsu National Synergistic Innovation Center for Advanced Materials, Nanjing Tech University, 211800 Nanjing, China.*

[3]*Department of Physics, Chemistry and Biology, Linkoping University, SE-581 83 Linkoping, Sweden.* E-mail: mats.fahlman@liu.se

[4]*CiC nanoGUNE Consolider, Tolosa Hiribidea, 76, E-20018 Donostia-San Sebastian, Spain*

[5]*CNR-ISMN, Via Gobetti 101, 40129 Bologna, Italy*

[6]*School of Physics and Optoelectronics, South China University of Technology, 510641 Guangzhou, China*

**ABSTRACT:**

Spinterface engineering has shown quite important roles in organic spintronics as it can improve spin injection or extraction. In this study, 11,11,12,12-tetracyanonaptho-2,6-quinodimethane (TNAP) is introduced as an interfacial layer for a prototype interface of Fe/TNAP. We report an element-specific investigation of the electronic and magnetic structures of Fe/TNAP system by use of near edge X-Ray absorption fine structure (NEXAFS) and X-ray magnetic circular dichroism (XMCD). Strong hybridization between TNAP and Fe and induced magnetization of N atoms in TNAP molecule are observed. XMCD sum rule analysis demonstrates that the adsorption of TNAP reduces the spin moment of Fe by 12%. In addition, induced magnetization in N K-edge of TNAP has also been found with other commonly used ferromagnets in organic spintronics, such as $La_{0.7}Sr_{0.3}MnO_3$ and permalloy, which makes TNAP a very promising molecule for spinterface engineering in organic spintronics.

# 1. Introduction

Spintronics, a branch of electronics, can take advantage of both the charge and the spin of the electron. And it involves the generation of a non-equilibrium spin polarization in various materials and devices, as well as its manipulation and detection. Among the most fascinating examples of spintronic applications are giant magnetoresistance (GMR) and tunneling magneto resistance (TMR). Recently, with the development of spintronics, several novel concepts and devices have also been exploited for different applications, for example, the realization of optically tunable MR promises the integration of the ultralow-power high-speed data writing and inter-chip communication[1]; graphene-based spintronic device shows promising solution for low-power devices beyond CMOS (complementary metal-oxide semiconductor) by manipulating the pure spin current[2]. In addition, spins can act as mediators to interconvert electricity, light, sound, vibration and heat, thus spin conversion can be realized, such as spin hall effects, spin Seebeck effect and spin Peltier effect [3].

In the past few decades, organic semiconductors (OSCs) have caught the attention of the spintronic community, and significant efforts are being made towards their integration in this field. Very recently, a molecular spin-photovoltaic device based on $C_{60}$ has been developed to integrate a photovoltaic response with the spin transport across the molecular layer, thus the photovoltaic response can be modified by the magnetic field[4]. Organic spintronic represents a very new and fascinating research field, where OSCs are used to mediate or control the spin polarized signal, because they consist mainly of atoms with low atomic number Z, leading to a low spin-orbital coupling and thus to extremely long spin relaxation time [5-7]. Efficient spin injection from a ferromagnetic (FM) electrode into an OSC is considered to be very important in organic spintronics [6]. FM/OSCs interfaces have recently become the subject of many studies, and spin injection/spin filtering at such interfaces is increasingly well understood [6-13].

In traditional organic electronics, the carrier barrier height at the metal-organic interface can be tuned by variety of buffer layers or carrier injection layers, and then carrier injection or extraction at metal-organic interface can be adjusted to improve device performance [14,15]. But in organic spintronics, the situation becomes more complicated. The reduction of carrier injection barrier is not enough because the interface is still spin undiscriminated, and then both charge carriers and their "spin" selection should be considered to improve "spin" injection. To get spin discriminated interface, it's important to generate different injection barrier for spin up and spin down at FM/OSCs interface (spin injector), or we can say spin-split bands. In this case, the spin discriminated interface acts as a spin filter to realize spin injection to the OSC layer. Hybrid interface states (HIS) are reported to be a feasible way to realize spin splitting at FM/OSCs interfaces, which may result from the chemical interaction between FM and OSC itself or additional interfacial layer [16,17]. HIS is considered to offer a strategy to engineer the spin injection in combination with appropriate molecules as the second layer by resonant tunneling [16]. In addition, it's also possible to tune OSCs/FM interface (spin collector) to realize net spin extraction by use of an interfacial layer [18]. In this case, the energy level shift at spin collector interface obtained in the interfacial layer may change the relevant spin band. As the spin-dependent hole extraction probability depends on the spin polarization in the cathode at the OSC highest occupied molecular orbital (HOMO) energy, a shift in the HOMO would change the spin polarization of extracted holes, which make it possible to select spin-up or spin-down to extract [18].

As well known, for current status in the community of organic spintronics, it's difficult to directly compare the results from different experiments because of the problem of common metrology rules. One of the most frequent discussion topics is related to the reproducibility of the published experimental results, and another is the discrepancy among the values reported by different groups as well as the relationship between MR and device resistance [6]. The interfacial effects are considered as the key factor which determines the spin injection, MR values and even the polarity (negative or positive) [6-9,18]. For example, for injection devices based on tris(8-hydroxyquinoline) aluminum ($Alq_3$) and Co top electrode, $La_{0.7}Sr_{0.3}MnO_3$ (LSMO) is widely used as the bottom electrode, and the MR shows negative [19], while it is positive when Fe is selected as the bottom electrode [20]. This comes from the different interfacial effects between Fe/$Alq_3$ and LSMO/$Alq_3$ and is assumed to be due to different hybridization induced spinterface conditions [9]. In addition, it has been shown that spinterface effects can be used to tune and modify the magnetic properties of the surface part of spin injecting electrodes through the hybridization with the first molecular layer [8, 21].

Recently, 11,11,12,12-tetracyanonaptho-2,6-quinodimethane (TNAP) has been reported to form HIS with fresh Co surface near the Fermi edge ($E_F$), and there is a strong dipole (0.7 eV) at the interface. Thus, it can improve carrier (hole) injection from Co through TNAP to OSCs as an interfacial layer. In addition, the induced magnetization in N K-edge has been found at Co/TNAP interface, which means that the spin polarization of injected carriers can be maintained, and then the spin injection can be assured from Co through TNAP to OSCs. In this case, TNAP shows very promising application for spinterface engineering by inserted between FM and OSCs in organic spintronics [17]. In this report, we will mainly focus on the prototype interface of Fe/TNAP. In fact,

Fe is a good candidate for spinterface characterization, since it has been popularly used in practical organic spintronic devices and confirmed to have an interesting interfacial effect with $Alq_3$ [13]. Here, we choose Fe as the substrate, one reason is that Fe has no near edge X-ray absorption fine structure (NEXAFS) features around N K-edge comparing with Co, in which the Co $L_2$ absorption peak will disturb the identification of the real signal from N K-edge, and then it will help to confirm the NEXAFS features in N K-edge. Another reason is that Fe has a higher chemical reactivity than Co, and we expect it may have much stronger hybridization at the interface, which will help to have clearer induced magnetization in TNAP molecules.

Here, we characterize the Fe/TNAP hybrid interface by use of NEXAFS and X-ray magnetic circular dichroism (XMCD). A strong hybridization between TNAP and Fe and induced magnetization in TNAP molecule are observed from the experiments. XMCD sum rule analysis demonstrates that the adsorption of TNAP reduces the spin moment of Fe by 12%, which is much stronger than that in the case of Co as the substrate. This comes from the different interfacial effects between Fe/TNAP and Co/TNAP due to the different hybridizations induced spinterface conditions resulted from the different chemical reactivity of Co and Fe. The results highlight the importance of choice of the ferromagnetic electrode in organic spintronics (which may differ between molecules). To confirm the induced magnetization in FM-TNAP interfaces, we have also discussed the results on $Fe_{80}Ni_{20}$ (permalloy, Py) and LSMO, which are both popularly used FM electrodes in organic spintronics [22-25].

**2. Experimental Section**

Experiments were performed at Linköping University (UPS/XPS) and D1011 beamline at the MAX-lab Synchrotron Facility in Lund (NEXAFS, XMCD).

*Sample preparation*: The TNAP was purchased from Tokyo Chemical Indusctry (TCI) Europe, which was used without purification. The base pressure in the experimental station is $2\times10^{-10}$ mbar, rising to $4\times10^{-10}$ mbar during Fe deposition, and $5\times10^{-9}$ mbar during TNAP deposition. For Fe thin film, Au-coated (200 nm) Si wafers was used as the substrates, and cleaned by Ar sputtering for 20 minutes before the metal deposition, which was confirmed by UPS and XPS. Fe thin film was deposited on clean Au surface by use of an UHV e-beam evaporator (Omicron EFM3) at a deposition rate of about 3 Å / min. The thickness of Fe was about 5~6 nm. TNAP was then deposited to get monolayer (ML) (0.8 nm) on Fe surface by the way reported previously [17]. All thickness was carefully estimated from the attenuation of the core level signals of the bottom layer (Au 4f), and all samples are in-situ fabricated and immediately transferred to analysis chamber for photoelectron spectroscopy.

Py pellets were purchased from Lesker with 99.95% purity. The thin films (d=10-20 nm) have been grown by means of electron-beam deposition in an ultra-high vacuum chamber ($P_{base}$= $5\times10^{-11}$ mbar) on $SiO_2$ (150 nm)/Si substrates [22]. The evaporation rate was set to 1.0 Å/s at a pressure of $5\times10^{-8}$ mbar, while the substrate temperature was hold at 25 °C during the deposition. Py samples were prepared in NanoGUNE, Spain, and were *in-situ* cleaned by Ar sputtering before experiments.

LSMO thin films were deposited by a homemade pulsed electron beam deposition setup in the channel spark ablation configuration on $NdGaO_3$ (110) (NGO) substrates at a pulse frequency of 6 Hz and a rate about 0.1 Å/pulse [24]; the substrate temperature was around 870 °C as measured by an optical pyrometer and the oxygen pressure in the chamber during the deposition was $3.5\text{-}4.0\times10^{-2}$ mbar. LSMO samples were prepared in ISMN-CNR, Bologna. After ultrasonic treatment in organic solvents, LSMO was heated in the solution so-called SC1 (5 $H_2O$, 1 $NH_4OH$, and 1 $H_2O_2$) at 85 °C for 5 minutes before spectroscopy experiments [25].

*Photonelectron spectroscopy*: The XPS and UPS experiments were carried out using a Scienta ESCA 200 spectrometer. The vacuum system consists of an analysis chamber and a preparation chamber. XPS and UPS measurements were performed in the analysis chamber at a base pressure of $10^{-10}$ mbar, using monochromatized Al ($K_\alpha$) X-rays at hv=1486.6 eV and He I radiation at hv=21.2 eV, respectively. The experimental conditions were such that the full width at half maximum (FWHM) of the Au 4f7/2 line was 0.65 eV. The binding energies were obtained referenced to the Fermi level with an error of ±0.1 eV. Sputtering and material depositions were done in a preparation chamber with a base pressure of $10^{-9}$ mbar.

*Synchrotron Experiments*: Both NEXAFS and XMCD were measured at room temperature. XMCD spectra were obtained in remanence, by taking the difference between NEXAFS spectra recorded with opposite in-plane magnetization directions. While for XMCD measurements on LSMO samples, the direction of light polarization is changed and the magnetization is kept. The samples were magnetized by applying an in-plane magnetic field pulse of 300 Oe. The angle of incidence of the photon beam was set to 40° relative to the sample surface, and the degree of circular polarization is 85% if there is no special description. All NEXAFS and XMCD measurements were normalized to the incident photon flux using the total electron yield (TEY) of a gold grid, on which a fresh layer of gold was deposited prior to the measurements.

## 3. Results and Discussions

As an electron acceptor, TNAP has shown strong interaction with clean metal surfaces forming hybrid interfaces [26], and it shifts the energy level alignment of metal-TNAP interface with a very strong dipole (**Figure S1** and **S2**). It has been reported to reduce hole injection barrier in real devices as it can enhance the work function of the anode [17,27]. In this report, we will mainly focus on the electronic and magnetic structures of a prototype interface for organic spintronics: Fe/TNAP, and also we will discuss the situation for the interfaces of Py/TNAP and LSMO/TNAP.

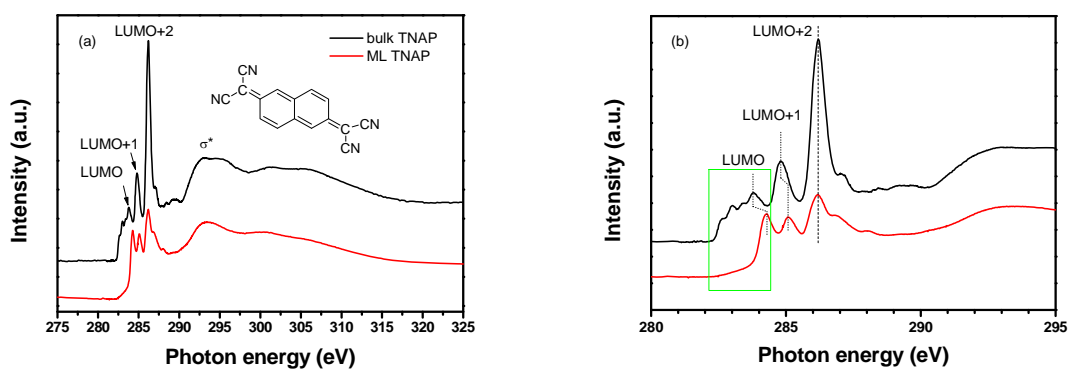

**Figure 1.** (a) C K-edge NEXAFS spectra for Fe/TNAP interface, and the inserted gives the molecular structure of TNAP. (b) Low photon energy region in (a), and the green rectangle indicates the frontier orbital states.

**Figure 1** shows NEXAFS spectra for C K-edge in both bulk and ML TNAP on Fe, recorded with a photon line-width of about 0.1 eV, and the inserted in Figure 1(a) gives the molecular structure of TNAP. Clear absorption features can be observed in the near edge region. A theoretical calculation of NEXAFS spectra of the gas-state TNAP, based on density functional theory (DFT) method, has been done in our reports previously [17]. It can be seen that the experimental results of bulk TNAP agree well with the calculation for the characteristic features. The discrepancy of the relative intensity and the peak position are considered to arise from the difference between gas and condensed states (mainly intermolecular screening). The most interesting phenomenon is the big difference in frontier orbitals between bulk and ML TNAP on Fe substrate. Comparing with the bulk molecule, several small features in low photon energy region are smeared by substrate underneath and disappear in C *K*-edge for ML TNAP (Figure 1a). This phenomenon may result from the efficient electron transfer from the outer orbitals of Fe to fill the frontier unoccupied molecular orbitals in ML TNAP. The LUMO peak (283.78 eV) in bulk shifts to higher photon energy of 284.28 eV in ML with a net value of 0.5 eV and becomes broadening, and LUMO+1 shifts the photon energy (284.8 eV) and appears at 285.1 eV (ML) with a net value of 0.3 eV, while the photon energy for LUMO+2 (286.2 eV) has nearly no change; only the peak becomes broadening (Figure 1b). The LUMO-shifting and peak-broadening indicate a strong interfacial bonding between TNAP molecules and Fe [28]. For Co/TNAP interface, all LUMO peaks in C K-edge have nearly no shift (only 0.03 eV or so), and the interaction between Co and TNAP is considered to be comparatively weak with carbon atoms, and it is mainly bonded from the nitrogen to the Co-surface [17]. When comparing with the results of Co/TNAP interface, C K-edge NEXAFS spectra are smeared/broadened for Fe/TNAP interface, which indicates interaction between Fe and TNAP through carbon bonding to the Fe-surface.

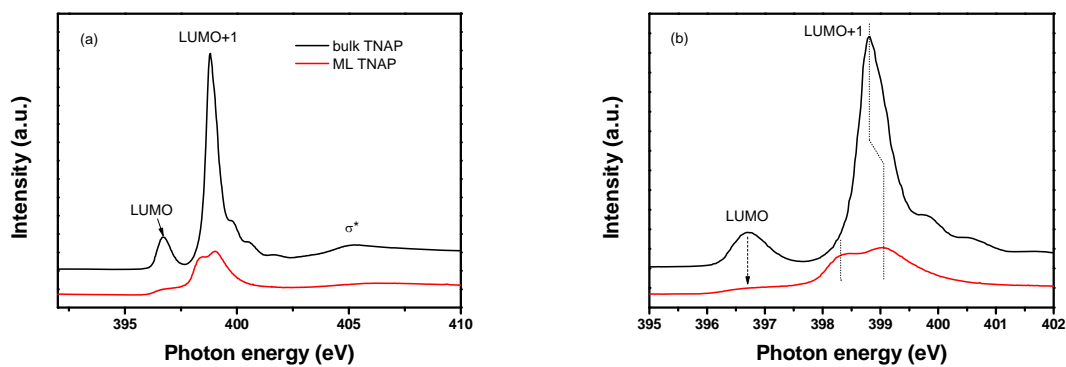

**Figure 2.** (a) N K-edge NEXAFS spectra for Fe/TNAP interface. (b) Low photon energy region in (a).

The spectra for N K-edge are much simpler than that for C K-edge in TNAP molecule, but similar hybridization-induced features have been also observed in N K-edge (**Figure 2**), and there are also peak-shifting and broadening in ML TNAP comparing with that in bulk. In Figure 2b, the LUMO peak (396.7 eV) becomes weaker and nearly disappears when ML TNAP is deposited on Fe surface, and the LUMO+1 peak slightly shifts to higher photon energy from 398.8 eV in bulk TNAP to 399.1 eV in ML TNAP with a net value of 0.3 eV. In addition, there is a new peak around 398.3 eV formed in ML TNAP on Fe, which is quite different from the results on Co/TNAP [17]. It suggests a stronger chemical interaction and orbital hybridization between Fe and ML TNAP than that between Co and ML TNAP through the bonding from N to Fe atoms on the surface.

UPS/XPS have also been carried out to determine the interaction/hybridization and bonding between Fe and TNAP (**Figure** S1, S2 and **S3**). There is a very strong interface dipole of 0.85 eV when ML TNAP is adsorbed on fresh Fe surface, and a HIS is clearly shown at 1.2 eV in the valence band from UPS. While for Co/TNAP, the interface dipole is 0.7 eV, and the HIS in valence band is much weaker than that for Fe/TNAP [17]. Thus, UPS results demonstrate a stronger interaction and hybridization at Fe/TNAP than that at Co/TNAP. As for XPS measurements, the main feature of C 1s is situated at 284.43 eV when a ML TNAP is adsorbed on Fe, and it gradually shifts to higher binding energy and becomes stable at 285 eV with the subsequent deposition of TNAP molecules until a bulk state is reached. The main feature around 285 eV in the bulk molecule becomes very wide with the range from 282.75 eV to 287.15 eV, which is quite different from that at Co/TNAP interface, and thus it's difficult to see the separated peak around 286.65 eV from the C in the cyano group. The C 1s for ML TNAP shows a much wider peak in Fe/TNAP than that in Co/TNAP. For N 1s spectra, a strong feature around 398.45 eV exists in ML TNAP on Fe, and its intensity is gradually reduced when more TNAP molecules are adsorbed on the Fe surface, and it finally disappears in the bulk TNAP. The main feature situated around 399.65 eV can be attributed to cyano groups as its intensity gradually becomes stronger with the increasing thickness. It should be indicated that the two separated peaks for N 1s are much clearer in ML TNAP on Fe than that on Co. Based on the discussion on XPS spectra of C 1s and N 1s, the chemical interaction between Fe and TNAP molecule happens mainly on N atoms in cyano groups, similarly to Co/TNAP, but it suggests a much stronger interaction in Fe/TNAP than for Co/TNAP.

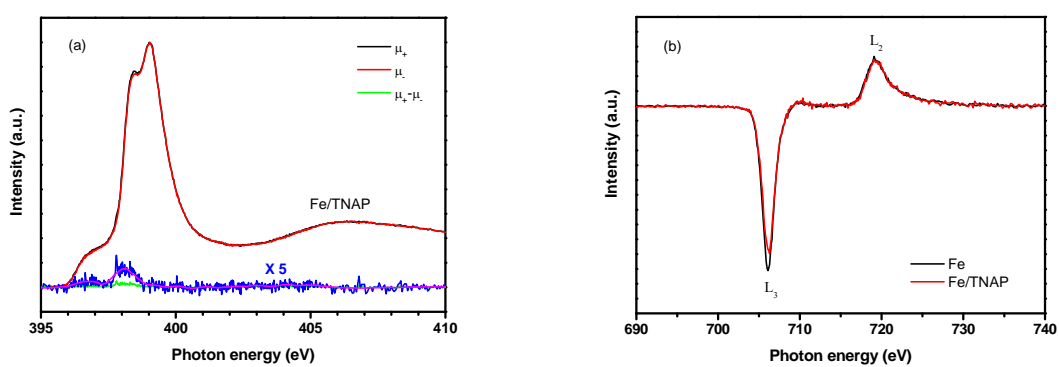

**Figure 3.** XMCD spectra in TEY mode for Fe/TNAP interface. (a) N K-edge. (b) Fe L-edge before and after adsorption of ML TNAP on Fe.

From above NEXAFS spectra for C and N K-edges, the interaction between TNAP and Fe has been considered to bond through both C and N atoms in cyano groups to Fe surface. Although C K-edge shows strong shifts for LUMO peaks, the N K-edge demonstrates significant difference between bulk and ML TNAP on Fe with a new feature around 398.3 eV at ML TNAP on Fe. And from XPS results, there is a new peak around 398.45 eV for N 1s in ML TNAP on Fe, while it's difficult to see the new feature from the interaction with C atoms. Furthermore, N atom has shown a stronger electronegativity than C atom [29]. Therefore, the interaction at Fe/TNAP may still preferentially happen between N and Fe atoms and partially between C and Fe atoms. In addition, as the cyano group is a strong acceptor, it's thus reasonable to conclude that mainly cyano groups participate in the chemical interaction at the hybrid interface. The different situations in NEXAFS and XPS for Co/TNAP and Fe/TNAP may be resulted from the stronger chemical reactivity of Fe than Co [29], based on which we can deduce a stronger interaction at Fe/TNAP than that at Co/TNAP.

NEXAFS and UPS/XPS results give the direct evidence of strong interaction and hybridization between TNAP molecule and Fe, and it's also interesting to know the induced magnetization in light elements of TNAP molecule by Fe electrode and counteraction effects on the magnetic property of Fe atoms by the molecules adsorbed [30]. **Figure 3** shows XMCD spectra for N K-edge (a) and Fe L-edge (b) in TEY mode for Fe/TNAP interface. The sizable XMCD signal of N K-edge further confirms the induced magnetization and spin polarization as in the previous report on the case of Co/TNAP [17]. There are two main features in XMCD spectra, one is situated at 396.8 eV, which belongs to the magnetization in LUMO state, and the other is situated at 398.4 eV, which belongs to the magnetization of the new peak state formed in ML TNAP on Fe. To our surprise, there is no XMCD feature at the peak position of 399.0 eV, which is considered to belong to the main feature in Co/TNAP interface, and it is also found in other interfaces such as Py/TNAP and LSMO/TNAP (**Figure S7**). The XMCD features in Fe/TNAP show much more different and stronger peaks than that in Co/TNAP (Figure 3a), which may result from the different interactions at Fe/TNAP and Co/TNAP, and this has been confirmed previously by NEXAFS and UPS/XPS measurements. In addition, it's also possible to find the different NEXAFS features resulted from the different interactions if we compare the XAS satellites of both samples of Co/TNAP and Fe/TNAP [31]. Indeed, Fe shows stronger interaction or hybridization with TNAP than Co and other substrates used here, such as Py and LSMO.

To investigate the induced magnetization in light elements of TNAP molecule at other hybrid interfaces, we choose two FM electrodes (Py and LSMO) which are popularly used in organic spintronics to carry out XMCD measurements (Figure S7). When Py is used as the substrate, the XMCD result looks quite similar to the one with Co in the shape without the feature from Co second order L-edge (Figure S7a). The main feature is just situated around 398.9 eV, and it shows a distinct peak with a small magnification of 5 times. Here, Py surface is sputtered clean by Ar ionized gas before the TNAP deposition, and then it's possible to have dangling bonds and clean atoms on the fresh surface which can have chemical interaction with cyano groups in TNAP molecule. From the valence band of sputtered Py substrate, there is a sharp Fermi edge, while it's flat near Fermi level in original Py without sputtering. A strong interface dipole of 0.9 eV and a distinct HIS at 1.1 eV are observed in UPS, which demonstrate a strong interaction and hybridization between TNAP and Py (**Figure S4** and **S5**). When LSMO/NGO is used as the FM substrate, it shows a very weak but clear feature for N K-edge in XMCD spectra by a magnification of 20 times (Figure S7b), and the peak position is exactly situated at 398.9 eV coming from the main feature of N K-edge in TNAP molecule, which is nearly the same as that on Py/TNAP and the reported Co/TNAP interfaces. The TNAP-LSMO interaction at the interface is likely aided by the cleaning process (SC1 method) that enhances the chemical reactivity of the LSMO surface [25].

Figure 3b shows the corresponding Fe L-edge XMCD spectra of the same sample before and after ML TNAP adsorption on Fe. The spectra are normalized to the $L_3$ peak height of the NEXAFS sum spectra for parallel and antiparallel alignment between the magnetization and photon helicity.

After adsorbing TNAP, the XMCD signal of Fe $L_3$-edge shows a clear decreased intensity while that of $L_2$-edge remains unchanged (within the experimental error), indicating a distinct reduction in the magnetic moments of Fe atoms due to the hybridization between Fe and TNAP electronic states. The hybridization was also observed at Co/TNAP interface, but the $L_3$-edge was slightly reduced in intensity after normalization of XMCD spectra [17], which may result from the different interaction at Fe/TNAP comparing with that at Co/TNAP interface due to the different chemical reactivity of Fe and Co.

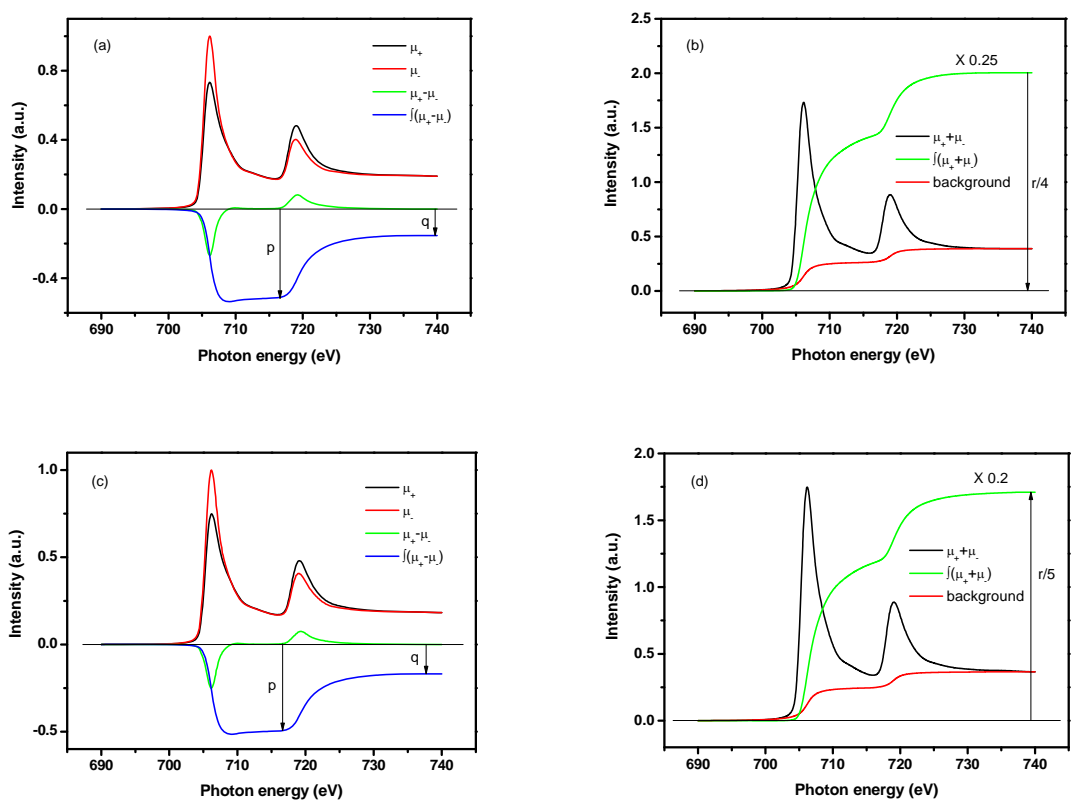

**Figure 4.** Sum rule analysis of Fe L-edge NEXAFS and XMCD spectra. (a) XMCD and (b) summed NEXAFS spectra and their integrations for clean Fe substrate, (c) XMCD and (d) summed NEXAFS spectra and their integrations for Fe substrate with ML TNAP adsorption.

XMCD sum rule analysis can deduce the element-specific orbital ($\mu_L$) and spin ($\mu_S$) magnetic moments from NEXAFS spectra and XMCD data, and it may provide possibility to know the change in the magnetization of Fe atoms before and after the molecule adsorption [30]. In **Figure 4**, the sum rule analysis of Fe L-edge XMCD and NEXAFS for Fe and Fe/TNAP (ML) are given. The details of the calculation of magnetic moments have been reported elsewhere [32]. Parameters p, q, and r are defined from the integrated areas of XMCD and NEXAFS sum spectra in Figure 4, and the electron occupation number ($n_{3d}$) for Fe atom is 6.61 based on the theoretical calculations [32]. For the sum rule analysis, the XMCD spectra are corrected by taking into account the incident angle (40° with respect to the sample surface) and the degree of circular polarization (85%), by multiplying the measured spectra by [1/cos(40°)]/0.85, while keeping the sum spectra unchanged. The calculation results of Fe magnetic moment are listed in **Table 1** for the same sample before and after TNAP adsorption. From the calculation, it shows a 12% reduction in spin moment ($\mu_s$) of Fe atoms with ML TNAP adsorption, but nearly no change in orbital moment ($\mu_L$). If comparing with TNAP on Co (0.8% reduction), the magnetic moment is reduced significantly, and this may be resulted from the stronger interaction between TNAP and Fe than Co. Because the interaction with the molecule overlayer is considered to be responsible for the reduction of FM magnetic moment at the interface [33,34], and the big degradation of magnetic moment in Fe may be due to the strong interaction between Fe and TNAP molecule. As Fe (1.83) and Co (1.88) have a larger difference in electronegativity in comparison to C (2.55) and N (3.04) [29], it's reasonable that there is stronger interaction at the interface of Fe/TNAP than that of Co/TNAP, and this has been confirmed by previous NEXAFS and UPS/XPS measurements. Based on above results and the demand of organic spintronics for FM electrodes, Fe probably is not a good candidate due to its distinct degradation in magnetic moments during the contact with molecules, while Co, LSMO and Py can be considered as FM electrodes for device application. These results thus may guide how to choose a suitable FM electrode for organic spintronics.

**Table 1.** Parameters and magnetic moments of Fe atoms before and after TNAP adsorption from XMCD sum rule analysis. (The values of p and q are calibrated by multiplying [1/cos(40°)]/0.85. )

|        | p      | q      | r     | $n_{3d}$ | $m_{spin}$ | $m_{orb}$ | $m_{orb}/m_{spin}$ |
|--------|--------|--------|-------|----------|------------|-----------|---------------------|
| Fe     | -0.789 | -0.235 | 8.017 | 6.61     | 1.604      | 0.133     | 0.083               |
| Fe/TNAP| -0.763 | -0.259 | 8.550 | 6.61     | 1.405      | 0.137     | 0.097               |

As we've discussed previously, spinterface is a very important aspect in organic spintronics, which may determine spin injection, MR values and even polarity (negative or positive) [6-9,18]. Our results demonstrate that the hybridization at FM-OSC interface can form HIS to improve spin injection, and may as well result in the counteraction on the magnetic property of FM electrodes. This may be one of reasons why there are different MR values and polarities during the measurements in organic spin valves with different FM electrodes [6,19,20].

## 4. Conclusion

In summary, we have investigated the *in-situ* electronic and magnetic structures of Fe/TNAP interface. As an electron acceptor, the adsorption of TNAP can form a big interface dipole on clean Fe surface and improve the work function up to 5.65 eV, which holds the promise to improve hole injection from Fe to organic semiconductor layer. The evolution of NEXAFS spectra of ML TNAP gives evidence of strong interaction between TNAP molecule and Fe, and the present results demonstrate the hybridization involves both C and N together, but mainly on N atoms in cyano groups. Induced magnetization in N atoms of TNAP molecule is observed by use of XMCD. Sum rule analysis demonstrates that the adsorption of TNAP molecule reduces the spin moment of Fe by 12%, which is much stronger than that in the case of Co (0.8%). These results sanction again the key importance of the spinterface properties for the realization of organic spintronic devices with required parameters. The spinterface is confirmed as a very complex and fundamental part of spintronic devices, defining the magnetic and spintronic properties at both sides of the hybrid interface. In addition, induced magnetization in N K-edge is also found with other ferromagnets in organic spintronics, such as LSMO and Py, which may provide the possibility of spinterface engineering with TNAP on such electrodes. As Py and LSMO are popularly used in organic spintronics, thus the results show promising application of TNAP to modify the spinterface in real spintronic devices.

**Acknowledgements**

This work was funded through the European Union Seventh Framework Programme (FP7/2007-2013) under grant agreement no. 263104 (Project HINTS). This work was also supported by the National Natural Science Foundation of China (Grant No. 61474046 , No. 51703173 and No. 11504168), and the Natural Science Foundation of Guangdong Province (Grant NO. 2017A030313).

# Supporting Information

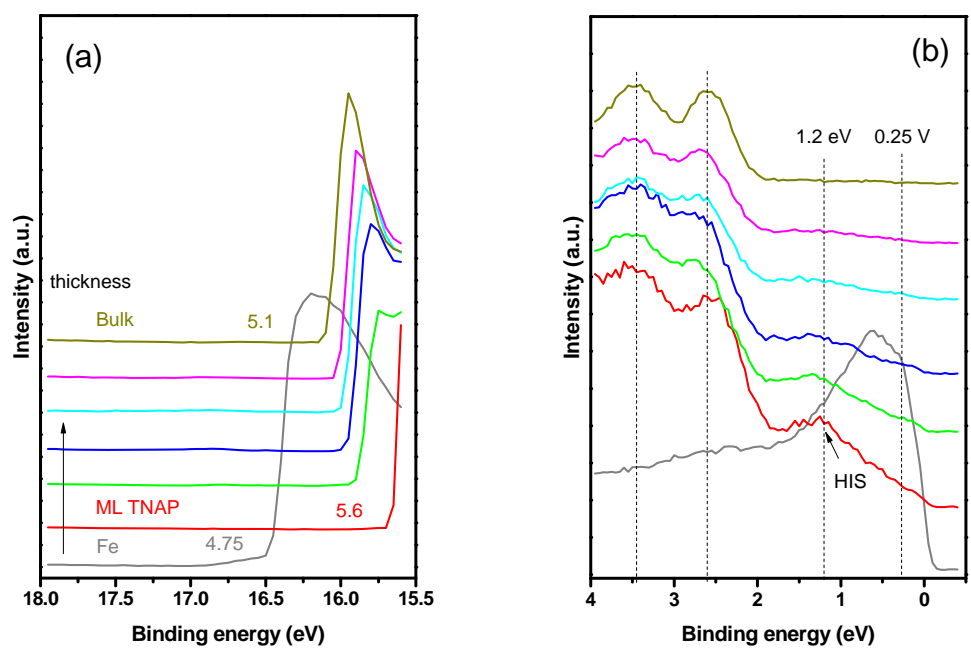

Figure S1. UPS of Fe/TNAP

Figure S1 gives the UPS measurements for Fe/TNAP interfaces with different thickness of TNAP, and the energy level alignment is shown in Figure S2. A strong interface dipole (0.85 eV) is formed when ML TNAP is deposited on Fe, and the work function is enhanced from 4.75 eV to 5.6 eV, which may help to reduce hole injection barrier with organic semiconductor (OSC) layer being deposited subsequently. A new peak around 1.2 eV appears at the valence band for ML TNAP (Figure S1a), and its intensity gradually decreases until it disappears with the thickness increasing. It hence can be assigned as HIS caused by the interaction between Fe and TNAP, similar to the case of TNAP on Co [1]. Because the peak around 0.25 eV below Fermi edge is mixed with the surface states from Fe, we can't assign it to another HIS as that in Co/TNAP interface. XPS measurements on C 1s and N 1s with the thickness increasing from ML to bulk molecule have been done to determine the interaction and bonding between Fe and TNAP (Figure S3), which have been discussed in the main text.

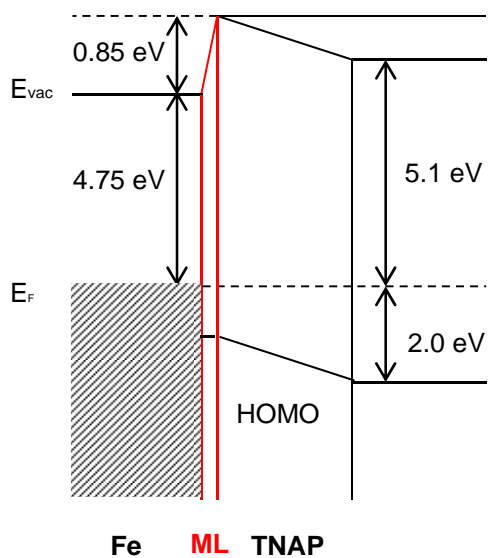

Figure S2. Energy level alignment of Fe/TNAP interface

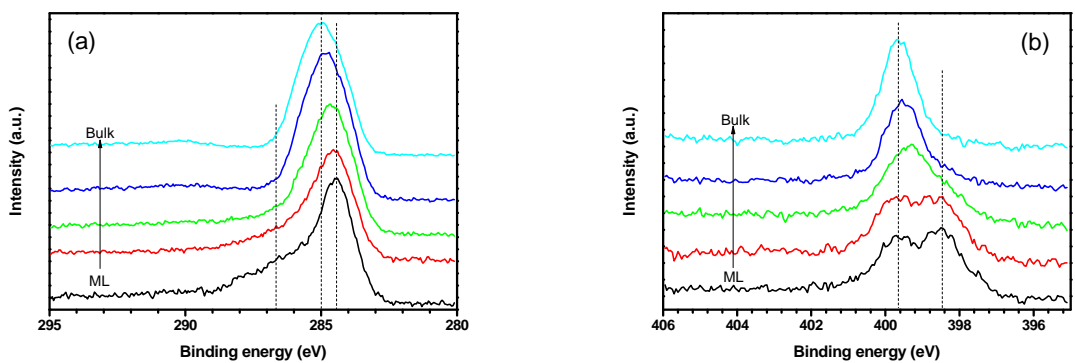

Figure S3. XPS for C1s and N 1s at Fe/TNAP

For Py substrate, we also find the similar evolution of work function with the thickness of TNAP by UPS. The work function of Py is enhanced from 4.75 eV to 5.65 eV upon the deposition of ML TNAP (Figure S4 and S5)[2]. A new HIS peak appears around 1.1 eV at the valence band at the interface, and also we can't assign the peak around 0.22 eV below the Fermi edge as the signal is mixed with Py surface states. In both cases of Fe/TNAP and Py/TNAP interfaces, the HIS shows a wide peak, which means a strong interaction between TNAP and such FM substrates [1].

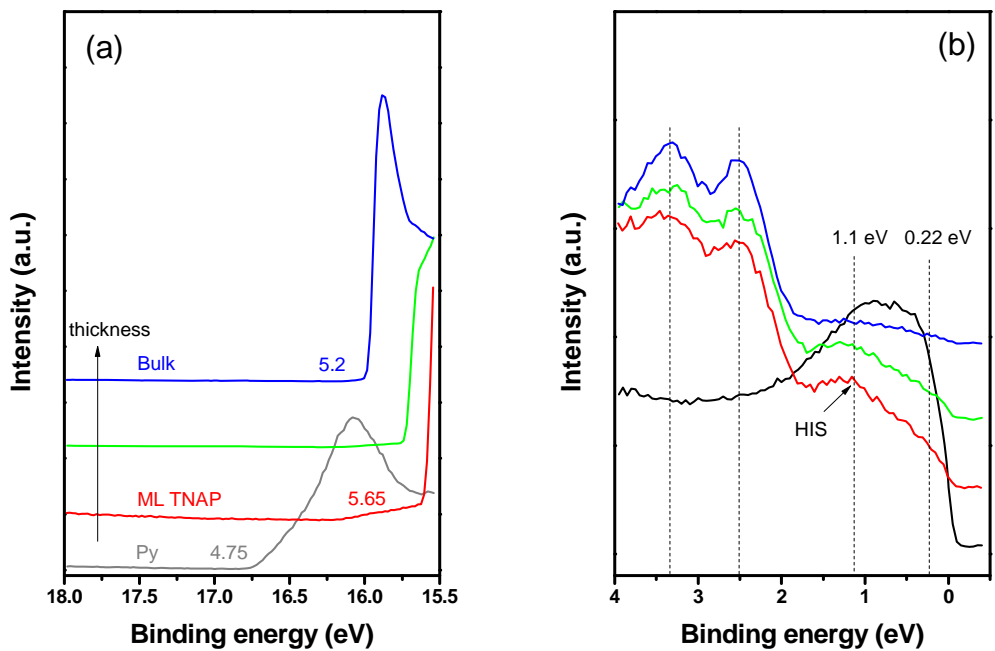

Figure S4. UPS of Py/TNAP interface

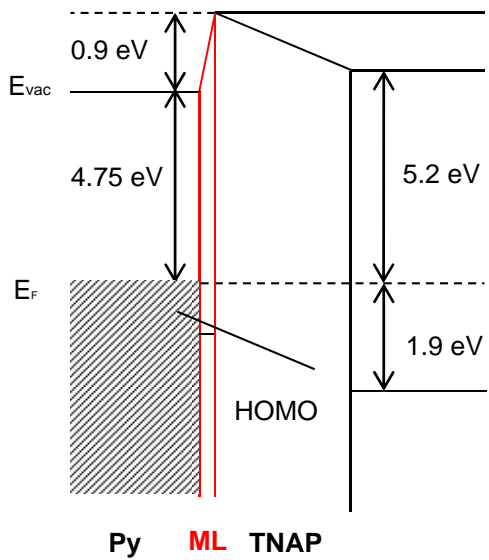

Figure S5. Energy level alignment of Py/TNAP interface

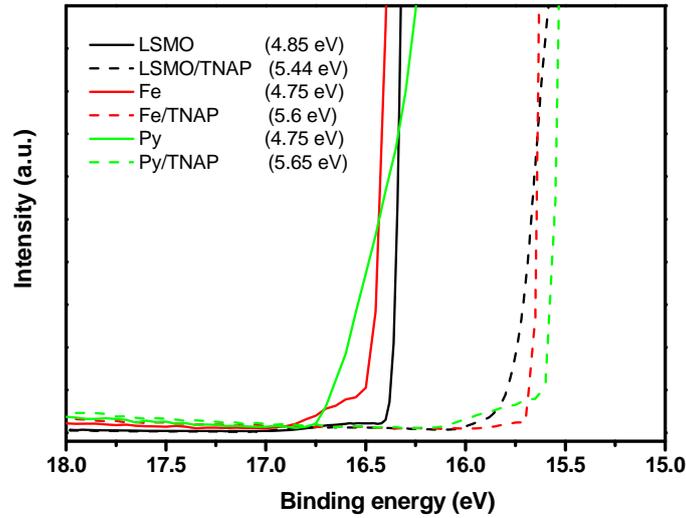

Figure S6. Comparison of secondary electron cutoffs for different ferromagnetic electrodes and corresponding interfaces with ML TNAP

LSMO is a popular FM electrode in organic spintronics for its high spin-polarization and stability [3]. Here, we have also checked the interface of LSO/TNAP by UPS. For comparison, we put together all secondary electron cutoffs in Figure S6 for different FM electrodes (Fe, Py, LSMO) and corresponding interfaces with ML TNAP, based on which work functions are calculated and indicated in the figure. We have found strong interface dipoles at both interfaces of Fe/TNAP (0.85 eV) and Py/TNAP (0.95 eV), and even at LSMO/TNAP (0.59 eV). From above, we may conclude that TNAP is a good interface layer to reduce carrier injection for FM electrodes.

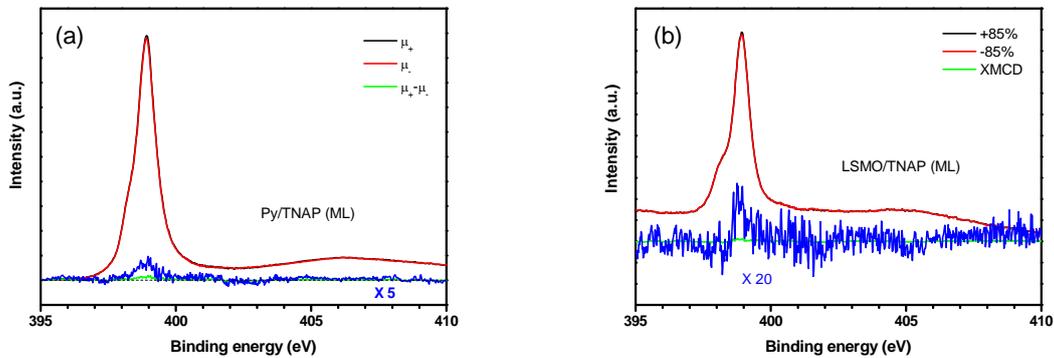

Figure S7. XMCD of Py/TNAP (a) and LSMO/TNAP (b)

Figure S7 gives XMCD for Py/TNAP and LSMO/TNAP interfaces, which are discussed in the main text.